\documentclass[12pt,eqsecnum]{revtex4}
\begin{document}
\title{Finite Nilpotent BRST transformations in Hamiltonian  formulation}


\author{ Sumit Kumar Rai \footnote{e-mail address: sumitssc@gmail.com}}
\author{Bhabani Prasad Mandal \footnote{e-mail address:
 \ \ bhabani.mandal@gmail.com, \ \  bhabani@bhu.ac.in  }}


\affiliation{ Department of Physics,\\
Banaras Hindu University,\\
Varanasi-221005, INDIA. \\
}

\begin{abstract}
We consider the finite field dependent BRST (FFBRST) transformations in the context of Hamiltonian formulation using Batalin-Fradkin-Vilkovisky method. The non-trivial Jacobian of such transformations is calculated in extended phase space. The contribution from Jacobian can be written as exponential of some local functional of fields which can be added to the effective Hamiltonian of the system. Thus, FFBRST in Hamiltonian formulation with extended phase space also connects different effective theories. We establish this result with the help of two explicit examples. We also show that the FFBRST transformations is similar to the canonical transformtaions in the sector of Lagrange multiplier and its corresponding momenta.
\end{abstract}
\maketitle
\newpage
\section{Introduction}
The Hamiltonian formulation developed by Batalin,Fradkin and Vilkovisky (BFV) \cite{frvi,frvi1,frvi2,frvi3} provides a powerful method for the Becchi-Rouet-Stora and Tyutin (BRST) \cite{brst}  quantization of systems with first class and second class constraints \cite{gara,nega,gaete,proca,chsh,rama1,rama2}. The important ingredients of BFV approach are as follows: the BRST transformations generated using BFV approach are independent of the gauge conditions, it does not rquire auxiliary field, it has an extended phase space where the ghosts and the Lagarange multipliers introduced act as dynamical variables. The nilpotent BRST charge in the BFV approach is directly constructed from the first class constraints which captures the algebraic structure of constraints in a gauge independent way whereas in the Lagarangian formulation the BRST charge is constructed using Noether's theorem from a gauge fixed Lagarangian.

BRST symmetry plays an important role in the quantization of gauge theories. BRST symmetry has been generalized in many ways such as non-local and non-covariant BRST \cite{lama}, covariant and non-local BRST \cite{tafi}, non-covariant and local BRST symmetry \cite{yale} and another local, covariant and off-shell nilpotent BRST symmetry \cite{rive,lahi}. They have been studied both in Lagrangian and Hamiltonian framework. In all the above generalizations of BRST symmetry, the parameter involved is infinitesimal, anticommuting and global. Furthermore, Joglekar and Mandal generalized the BRST symmetry transformations by taking the parameter to be finite field-dependent instead of infinitesimal but anticommuting and global \cite{jm}. This generalized BRST, so called the finite field dependent BRST (FFBRST) transformations are also the symmetry of the effective action but the path integral measure changes in a non-trivial way. It has found many applications \cite{sdj0,cou,sdj,rb,sdj1,etc,rama,upma,mara}. For example, the gauge field propagators in non-covariant gauges contain singularities on the real momentum axis. Proper prescriptions for these singularities in gauge field propagators has been found by connecting theory in Lorentz gauge to theory in axial gauge by using generalized BRST transformations \cite{sdj0}. It has also been used to regularize the divergence in energy integral in Coulomb gauge  \cite{cou}. 

It has been already shown that finite BRST connects generating functionals corresponding to solutions of quantum master equations in field/antifield formulation of pure YM theory \cite{rama}. Such generalized BRST symmetry is also the formal symmetry of the generating functional in field/antifield formulation of pure YM theory \cite{mara}. So far FFBRST formulation and all its applications are considered only in the Lagrangian framework. It will be interesting to extend FFBRST formulation in the Hamiltonian framework. In particular, FFBRST in Hamiltonian formulation will be useful for  Sp(2) and Sp(3)  BRST quantization in extended phase space   \cite{bala1,bala2,coio,lamo,coio1,roro,abbr}. In order to do so, the formulation of finite BRST in Hamiltonian framework is must.

In this paper, we extend FFBRST formulation  in 
Hamiltonian framework and show that how the non-trivial Jacobian arising in the path integral measure plays the role in the extended phase space using BFV formulation. We show that the non-trivial Jacobian of the path integral measure can always be expressed as $e^{iS_1}$, where $S_1$ is some local function of 
field variables in the extended phase space and can be a part of the effective action. Thus, FFBRST can connect the 
generating functionals of two different effective field theories with suitable choice of  
finite field dependent parameter in the BFV formulation.
For example, we show that FP
effective action in Lorentz gauge with a gauge parameter $\lambda $ can be connected to
(i) FP effective action  in Coulomb gauge ,
(iv) FP effective action  with another distinct gauge parameter 
$\lambda^\prime $ in the BFV Hamiltonian formulation.
 Exploring the advantages of BFV Hamiltonian framework, we also show that the finite field dependent BRST transformations is canonical transformation in the case of Coulomb gauge.
 
The plan of our paper is as follows. In Sec. II, we discuss the important features of BFV formulation. In Sec. III, we discuss the FFBRST transformations in the extended phase space for a general framework where we calculate the Jacobian change. We have considered two examples. In Sec. IV, we show that FFBRST transformtaion is nothing but canonical in the case of Coulomb gauge. We conclude in Sec. V .
 
\section{BFV formalism}
BFV formalism is extremely useful in the quantization of syatems with first class constraints. We present here only the essence of this approach and the detailed discussion can be found in \cite{frvi,frvi1,frvi2,frvi3}. The action in finite phase space can be expressed as 
\begin{equation}
S=\int dt \left( p^\mu\dot{q}_\mu - H_c -\lambda^a \Omega_a \right),
\end{equation}
where ($q^\mu,p_\mu$) are the canonical variables describing the theory. $H_c$ is the 
canonical Hamiltonian and $\lambda^a$ are the Lagrange multiplier associated with first class 
constraints, $\Omega_a$. In this approach, Lagrange multipliers $\lambda^a$ and its corresponding conjugate canonical momenta $p_\lambda^a$ are dynamical 
variables. The
canonical momenta  to $\lambda^a$ i.e. $p_\lambda^a$ must be imposed as new constraints 
such that the dynamics of the theory does not change.
BFV formalism extends the phase space by introducing two canonical pairs of ghost ( ${\cal{C}}^a,{\cal{P}}^a$) and ( ${\bar{\cal{C}}}^a,{\bar{\cal{P}}}^a$) for each of the constraints satisfying the following equal-time anticommutation relations
\begin{eqnarray}
\left \{{\cal{C}}^a({\bf x},t) ,{\cal{P}}^b ({\bf y},t)\right\} &=& -i\delta^{ab}\delta ({\bf x}-{\bf y}), \nonumber\\
\left \{{\bar{\cal{C}}}^a({\bf x},t) ,{\bar{\cal{P}}}^b ({\bf y},t)\right\} &=& -i\delta^{ab}\delta ({\bf x}-{\bf y}).\label{antic}
\end{eqnarray}.
In this formalism, the nilpotent generator for the system with first class constraint have the general form as 
\begin{equation}
Q={\cal{C}}_a \Omega^a + \frac{1}{2}{\cal{P}}^a f^{bc}_a {\cal{C}}_b {\cal{C}}_c+{\bar{\cal{P}}}^a p_\lambda^a, 
 \label{sg}
\end{equation}
where the $f^{bc}_a$ is a structure constant, $\Omega^a $ is the first class constraints.
According to Fradkin-Vilkovisky \cite{frvi} theorem, the generating functional in the extended phase space is given by
\begin{equation}
Z_\Psi =\int {\cal{D}}\varphi \; \exp (iS_{eff}) \label{zpsi},
\end{equation}
where the effective action, $S_{eff}$ is 
\begin{equation}
S_{eff}=\int dt\left (p^\mu {\dot q}_\mu + {\dot{\cal{C}}}^a {\cal{P}}_a + p^a\dot{\lambda}_a 
-H_\Psi \right ). \label{seff}
\end{equation}
${\cal{D}}\varphi$ is the Liouville measure on the phase space. $H_\Psi$ is the extended 
hamiltonian given as
\begin{equation}
H_\Psi =H_c + \left\{Q,\Psi\right\}.
\end{equation}
$\Psi$ is the gauge fixed fermion and $Z_\Psi$ does not depend upon the choice of $\Psi$ \cite{henn}.

\section{Finite BRST in BFV formulation}
The BRST transformations are generated from the charge given by Eq. (\ref{sg}) using $\delta\varphi=\left[\varphi, Q\right]\delta \lambda$ where $\delta\lambda$ is infintesimal anticommuting BRST parameter under which the effective action in Eq. (\ref{seff}) remains invariant. Joglekar and Mandal generalized the anticommuting BRST parameter $\delta\lambda$ to be finite field-dependent instead of infinitesimal but  space time independent parameter $\Theta[\phi]$. Under this generalization, the path integral measure ${\cal{D}}\phi $ will give rise to a Jacobian J. The Jacobian for this finite BRST transformations for certain $\Theta[\phi]$ can be calculated by writing the Jacobian as a succession of infinitesimal transformations. We can write 
\begin{eqnarray}
{\cal{D}}\phi&=&J(\kappa){\cal{D}}\phi^\prime(\kappa)\nonumber\\
&=&J(\kappa+d\kappa){\cal{D}}\phi^\prime(\kappa+d\kappa)\label{jacoeva}
\end{eqnarray}

We have introduced a numerical parameter  $\kappa:0\leq\kappa\leq 1$. All the fields are taken to be the function of $\kappa$. For a generic field $\phi(x,\kappa)$, $\phi(x,0)=\phi(x)$ and $ \phi(x,\kappa=1)=\phi^\prime$.
Now the transformations from $\phi(\kappa)$to $\phi(\kappa+d\kappa)$ is an infinitesimal one, therefore Jacobian can be expressed as 
\begin{eqnarray}
\frac{J(\kappa)}{J(\kappa+d\kappa)}&=&\sum_\phi\pm\frac{\delta\phi(x,\kappa+d\kappa)}{\delta\phi(x,\kappa)},\nonumber\\
&=&1-\frac{1}{J(\kappa)}\frac{dJ(\kappa)}{d\kappa}d\kappa
\label{jaco}
\end{eqnarray}
$\sum_\phi$ is sum over all the fields in extended phase space 
$p_\mu,q_\mu,p_\lambda,\lambda,{\cal{P}}^a,{\bar{\cal{P}}}^a,{\cal{C}}^a,{\bar{\cal{C}}}^a$ (as in Eq. (\ref{seff})) and $\pm$ sign for whether $\phi$ is bosonic or fermionic.

Now we consider the generating functional
\begin{equation}
{\cal{Z}}=\int {\cal{D}}\phi(x,0)\; e^{iS_{eff}\left[\phi(x,0)\right]},\label{zlj}
\end{equation}
where $\phi(x,0)$ generically denotes all the fields in extended phase space at $\kappa=0$. This generating functional is equal to
\begin{eqnarray}
&&\int {\cal{D}}\phi(x,\kappa)\;J(\kappa)\; e^{iS_{eff}\left[\phi(x,\kappa)\right]},\nonumber\\
&=&\int {\cal{D}}\phi(x,\kappa+d\kappa)\;J(\kappa+d\kappa)\; e^{iS_{eff}\left[\phi(x,\kappa+d\kappa)\right]}\nonumber\\
&=&\int {\cal{D}}\phi(x,\kappa+d\kappa)\;J(\kappa)\left[1+\frac{1}{J}\frac{dJ}{d\kappa}d\kappa\right]\; e^{iS_{eff}\left[\phi(x,\kappa+d\kappa)\right]}
\end{eqnarray}

where the invariance of the $S_{eff}$ under $\phi(x,0)\;\rightarrow\phi(x,\kappa)$ is a BRST transformations given by 
\begin{equation}
\phi(0)=\phi(\kappa)-\delta_b\phi(\kappa)\;\Theta\left[\phi(\kappa),\kappa\right]
\end{equation}
where $\phi^\prime=\phi(\kappa=1)$ and $\phi=\phi(\kappa=0).$ 
$J(\kappa$ can be replaced by $e^{iS_1\left[\phi(\kappa);\kappa\right]}$ for a certain functional $S_1$ which needs to be determined in each individual case if and only if the following condition is satisfied \cite{jm}
\begin{equation}
\int{\cal{D}}\phi(\kappa)\left[\frac{1}{J}\frac{dJ}{d\kappa}-i\frac{dS_1}{d\kappa}\right]e^{i\left(S_1+S_{eff}\right)}=0,\label{jc}
\end{equation}
where $\frac{dS_1}{d\kappa}$ is a total derivative of $S_1$ with respect to $\kappa$ in which dependence on $\phi(\kappa)$ is also differentiated and  the Jacobian  can be expressed as  $e^{iS_1}$ where $S_1$ is some local functionals of fields and it satisfies the following condition \cite{jm}
\begin{equation}
\frac{1}{J}\frac{dJ}{d\kappa}-\frac{dS_1}{d\kappa}=0. \label{expco}
\end{equation}

\vspace{0.2in}
{\bf \large An example: YM theory}

We consider pure Yang-Mills theory in this formulation to construct the symmetry generator, effective action and finally the generating functional using BFV formulation in extended phase space.
The kinetic part of the Yang-Mills Lagrangian is given by
\begin{equation}
{\cal{L}}=-\frac{1}{4}F_{\mu\nu}^a F^{\mu\nu a},\label{kin}
\end{equation}
where, $F_{\mu\nu}^a=\partial_\mu A_\nu^a-\partial_\nu A_\mu^a +ig f^{abc}A_\mu^b A_\nu^c.$
The canonical momenta corresponding to $A_0$ and $A_i$ fields are
\begin{eqnarray}
\Pi_0^a &=&\frac{\partial{\cal{L}}}{\partial{{\dot{A}}_0^a}}=0, \label {pc}\\
\Pi_i^a &=&\frac{\partial{\cal{L}}}{\partial{{\dot{A}}_i^a}}=F_{0i}^a.
\end{eqnarray}
The canonical Hamiltonian density becomes
\begin{equation}
{\cal{H}}_c=\frac{1}{2}\Pi_i^a\Pi_i^a+\frac{1}{4}F_{ij}^aF_{ij}^a-A_0^aD_{ab}^i\Pi_i^b. \label{canh}
\end{equation}
$\Omega_1=\Pi_0^a$ is the primary constraint of the theory. Using Dirac prescriptions \cite{dirac}, we can find the the secondary constraints from  primary constraints $\Omega_1$ as 
\begin{equation}
\Omega_2={\dot{\Omega}}_1=\left\{H_c,\Omega_1\right\}=D^i_{ab}\Pi_i^a\approx0, \label{sc}
\end{equation}
where, $D_i^{ab}=\partial_i\delta^{ab}+gf^{abc}A_i^c$ is the covariant derivative.
These two constraints form a set of first class constraints as they satisfy a Poisson bracket 
\begin{equation}
\left\{\Omega^a,\Omega^b\right\}=0.
\end{equation}

Using the BFV formalism, the BRST charge is  constructed using Eq. (\ref{sg}) for this theory as
\begin{equation}
Q={\cal{C}}_a D^i\Pi_i^a + \frac{1}{2}{\cal{P}}^a f^{bc}_a {\cal{C}}_b {\cal{C}}_c+{\bar{\cal{P}}}^a \Pi_0^a. 
 \label{brstc}
\end{equation}
The effective action in the extended phase space can be expressed as 
\begin{equation}
S_{eff}=\int d^4 x \left[\Pi^{ia}A_i^a+\Pi^{0a}A_0^a+{\dot{\cal{C}}}^a{\cal{P}}^a + {\dot{\bar{\cal{C}}}}^a{\bar{\cal{P}}}^a-{\cal{H}}_c-\left\{\Psi,Q\right\}\right],\label{effa}
\end{equation}
where, ${\cal{H}}_c$ is the canonical Hamiltonian density given by Eq. (\ref{canh}). $\Psi$ is the gauge fixed fermion which for the Lorentz gauge can be chosen as
\begin{equation}
\Psi=\int d^3x\left[\frac{\lambda}{2}\;{\bar{\cal{C}}}\Pi_0^a+{\bar{\cal{C}}}\partial_iA^{ia}+{\bar{\cal{P}}}^aA_0^a\right].\label{psil}
\end{equation}
Putting Eq. (\ref{brstc}) and Eq. (\ref{psil}) in Eq. (\ref{effa}), we obtain the effective action as 
\begin{eqnarray}
S_{eff}&=&\int d^4x\left[ \Pi^{ia}{\dot{A_i}}^a+\Pi^{0a}{\dot{A_0}}^a+{\cal{P}}^a{\dot{\cal{C}}}^a +{\bar{\cal{P}}}^a {\dot{\bar{\cal{C}}}}^a-\frac{1}{2}\Pi_i^a\Pi_i^a-\frac{1}{4}F_{ij}^aF_{ij}^a+A_0^aD_{ab}^i\Pi_i^b \right. \nonumber \\
&+&\left.\Pi_0^a\partial_iA^{ia}+\frac{\lambda}{2}\Pi_0^a\Pi^{0a}+
{\bar{\cal{P}}}^a{\cal{P}}^a-\partial_i{\bar{\cal{C}}}^aD^i{\cal{C}}^a+gf^{abc}{\cal{P}}^a{\cal{C}}^bA_0^c \right].\label{fiac}
\end{eqnarray}

 The effective action given in Eq. (\ref{fiac}) is invariant under the set of following BRST transformations with a infinitesimal parameter generated by the BRST charge in Eq. (\ref{brstc}):
\begin{eqnarray}
\delta_bA_0&=&{\bar{\cal{P}}}^a\;\delta\Lambda,\quad\quad\quad\quad\quad\quad\quad\quad\quad\quad\quad\delta_bA_i=D_i{\cal{C}}^a\;\delta\Lambda, \nonumber\\
\delta_b {\cal{C}}^a_i&=&\frac{1}{2}f^{abc}{\cal{C}}_b{\cal{C}}_c\;\delta\Lambda,\quad\quad\quad\quad\quad\quad\quad\quad\delta_b{\bar{\cal{C}}}^a=\Pi_0^a\;\delta\Lambda,\nonumber\\
\delta_b{\cal{P}}^a&=& \left(D_i^{ab}\Pi^{ia}+gf^{abc}{\cal{P}}^b{\cal{C}}^c\right)\delta\Lambda,\quad\quad\quad \delta_b {\bar{\cal{P}}}=0,\nonumber\\
\delta_b\Pi_0^a&=&0,\quad\quad\quad\quad\quad \quad\quad\quad\quad\quad\quad\quad\;\;\quad\delta_b\Pi_i^a=0, \label{usubrs}
\end{eqnarray}
where $\delta\Lambda$ is the global, infinitesimal and anticommuting parameter. Similar to the FFBRST formulation in the framework of Lagrangian formulation, we can construct a finite version of the BRST transformations in BFV approach as
\begin{eqnarray}
\delta_bA_0&=&{\bar{\cal{P}}}^a\;\Theta,\quad\quad\quad\quad\quad\quad\quad\quad\quad\quad\quad\delta_bA_i=D_i{\cal{C}}^a\;\Theta, \nonumber\\
\delta_b {\cal{C}}^a_i&=&\frac{1}{2}f^{abc}{\cal{C}}_b{\cal{C}}_c\;\Theta,\quad\quad\quad\quad\quad\quad\quad\quad\delta_b{\bar{\cal{C}}}^a=\Pi_0^a\;\Theta,\nonumber\\
\delta_b{\cal{P}}^a&=& \left(D_i^{ab}\Pi^{ia}+gf^{abc}{\cal{P}}^b{\cal{C}}^c\right)\Theta,\quad\quad\quad \delta_b {\bar{\cal{P}}}=0,\nonumber\\
\delta_b\Pi_0^a&=&0,\quad\quad\quad\quad\quad \quad\quad\quad\quad\quad\quad\quad\;\;\quad\delta_b\Pi_i^a=0.\label{ffbrst}
\end{eqnarray}
$\Theta$ is the finite-field dependent, global and anticommuting Parameter i.e.$\left(\Theta^2=0\right)$. The transformations given in Eq. (\ref{ffbrst}) also leaves the effective action in Eq. (\ref{fiac}) invariant.
The generating functional in the BFV formalism can be expressed as 
\begin{eqnarray}
Z&=&\int dA_0^a\;dA_i^a \;d\Pi_0^a\;d\Pi_i^a\;d{\cal{P}}^a{\bar{\cal{P}}}^a\;d{\bar{\cal{C}}}^a\;d{\cal{C}}^a\exp({iS_{eff}}),\nonumber\\
&=&\int D\chi \exp\left[i\int d^4x\left\{\Pi^{ia}A_i^a+\Pi^{0a}A_0^a+{\cal{P}}^a{\dot{\cal{C}}}^a +{\bar{\cal{P}}}^a {\dot{\bar{\cal{C}}}}^a-\frac{1}{2}\Pi_i^a\Pi_i^a-\frac{1}{4}F_{ij}^aF_{ij}^a+A_0^aD_{ab}^i\Pi_i^b \right.\right. \nonumber \\
&+&\left.\left.\Pi_0^a\partial_iA^{ia}+\frac{\lambda}{2}\Pi_0^a\Pi^{0a}+
{\bar{\cal{P}}}^a{\cal{P}}^a-\partial_i{\bar{\cal{C}}}^aD^i{\cal{C}}^a+gf^{abc}{\cal{P}}^a{\cal{C}}^bA_0^c \right\}\right],\label{zfiac} \\
{\mbox where,}\nonumber \\ 
{\cal{D}}\phi&=&dA_0^a\;dA_i^a \;d\Pi_0^a\;d\Pi_i^a\;d{\cal{P}}^a{\bar{\cal{P}}}^a\;d{\bar{\cal{C}}}^a\;d{\cal{C}}^a.
\end{eqnarray}
${\cal{D}}\phi$ is the path integral measure which is integrated over all the phase space. The finite transformation given in Eq. (\ref{ffbrst}) leaves the effective action in Eq. (\ref{fiac}) invariant but the path integral measure ${\cal{D}}\phi$ in the generating functional Eq. (\ref{zfiac}) is not invariant. It gives rise to a Jacobian in the extended phase space and needs to be calculated.

The Jacobian using Eq. (\ref{jacoeva})  is calculated in the extended phase space as follows
\begin{eqnarray}
&&dA_0 dA_i d\Pi_0 d\Pi_i d{\cal{P}}d{\bar{\cal{P}}}d{\cal{C}}d{\bar{\cal{C}}} \nonumber\\
&=& J(\kappa)dA_0(\kappa) dA_i(\kappa) d\Pi_0(\kappa) d\Pi_i(\kappa) d{\cal{P}}(\kappa)d{\bar{\cal{P}}}(\kappa)d{\cal{C}}(\kappa)d{\bar{\cal{C}}}(\kappa)\nonumber\\
&=& J(\kappa+d\kappa)dA_0(\kappa+d\kappa) dA_i(\kappa+d \kappa) d\Pi_0(\kappa+d\kappa) d\Pi_i(\kappa+d\kappa) d{\cal{P}}(\kappa+d\kappa)d{\bar{\cal{P}}}(\kappa+d\kappa)\nonumber\\
&&d{\cal{C}}(\kappa+d\kappa)d{\bar{\cal{C}}}(\kappa+d\kappa).
\end{eqnarray} Expanding R.H.S of Eq. (\ref{jaco}), we obtain
\begin{eqnarray}
&&\int d^4x \sum_a\left[\frac{\delta A_0^a(x,\kappa+d\kappa)}{\delta A_0^a(x,\kappa)}+\frac{\delta A_i^a(x,\kappa+d\kappa)}{\delta A_i^a(x,\kappa)}- \frac{\delta {\cal{C}}^a(x,\kappa+d\kappa)}{\delta {\cal{C}}^a(x,\kappa)} -\frac{\delta {\bar{\cal{C}}}^a(x,\kappa+d\kappa)}{\delta {\bar{\cal{C}}}^a(x,\kappa)}\right. \nonumber\\
&& \left. - \frac{\delta {\cal{P}}^a(x,\kappa+d\kappa)}{\delta {\cal{P}}^a(x,\kappa)}-\frac{\delta {\bar{\cal{P}}}^a(x,\kappa + d\kappa)}{\delta {\bar{\cal{P}}}^a(x,\kappa)}+\frac{\delta \Pi_0^a(x,\kappa+d\kappa)}{\delta \Pi_0^a(x,\kappa)}+\frac{\delta \Pi_i^a(x,\kappa+d\kappa)}{\delta \Pi_i^a(x,\kappa)}\right].
\end{eqnarray}

This equation can be further expanded as 
\begin{eqnarray}
1&+&d\kappa\int\left[ {\bar{\cal{P}}}^a\frac{\delta\Theta^\prime[\phi(x,\kappa)]}{\delta A_0^a(x,\kappa)} +D_i^{ab}{\cal{C}})_b \frac{\delta\Theta^{\prime a}[\phi(x,\kappa)]}{\delta A_i^a(x,\kappa)}-\Pi_0^a(x,\kappa)\frac{\delta\Theta^\prime[\phi(x,\kappa)]}{\delta {\bar{\cal{C}}}^a(x,\kappa)}\right.\nonumber\\
&-&\left.gf^{abc}{\cal{C}}^b(x,\kappa){\cal{C}}^c(x,\kappa)\frac{\delta\Theta^\prime[\phi(x,\kappa)]}{\delta {\cal{C}}^a(x,\kappa)}-\left(D_i^{ab}\Pi^i_b+gf^{abc}{\cal{P}}^b{\cal{C}}^c\right)\frac{\delta\Theta^\prime[\phi(x,\kappa)]}{\delta {\cal{P}}^a(x,\kappa)}\right]\nonumber\\
&=&\frac{J(\kappa)}{J(\kappa+d\kappa)},\nonumber\\
&=&1-\frac{1}{J(\kappa)} \frac{dJ(\kappa)}{d\kappa}d\kappa.\label{jaco1}
\end{eqnarray}

{\bf Case I:}\\
For a choice of finite BRST parameter $\Theta$ related to $\Theta^\prime$ as follows
\begin{equation}
\Theta^\prime=\int d^4x\left[\gamma_1\lambda \;{\bar{\cal{C}}}^a\;\Pi_0^a+\gamma_2\;{\bar{\cal{C}}}^a \;\partial_0A_0^a\right],\label{tc}
\end{equation}
the Jacobian change can be calculated from  Eq. (\ref{jaco1}) as follows
\begin{equation}
\frac{1}{J}\frac{dJ}{d\kappa}=\int d^4x\left[\gamma_1\lambda\;{\Pi_0^a}^2+\gamma_2\;\Pi_0^a\partial_0A_0^a+\gamma_2\;{\bar{\cal{C}}}^a\partial_0{\bar{\cal{P}}}^a\right].
\end{equation}
We make an ansatz  for $S_1$ as follows
\begin{equation}
S_1=\int d^4x\left[\xi_1(\kappa){\Pi_0^a}^2+\xi_2(\kappa)\Pi_0^a\partial_0A_0+\xi_3(\kappa){\bar{\cal{C}}}^a\partial_0{\bar{\cal{P}}}^a\right],
\end{equation}
where, $\xi^{\prime s}_i$ are arbitrary constants which depend explicitly on $\kappa$ whereas the fields $\phi_i^{\prime s}$ have implicit $\kappa$ dependence. We calculate
\begin{eqnarray}
\frac{dS_1}{d\kappa}&=&\int d^4x \left[\xi_1^\prime(\kappa){\Pi_0^a}^2+\xi_2^\prime(\kappa)\Pi_0^a\partial_0A_0^a+\xi_3^\prime(\kappa){\bar{\cal{C}}}^a{\dot{\bar{\cal{P}}}}^a+\xi_2(\kappa)\Pi_0^a{\dot{\bar{\cal{P}}}}^a\Theta^\prime\right. \nonumber\\
&-&\left.\xi_3(\kappa)\Pi_0^a{\dot{\bar{\cal{P}}}}^a\Theta^\prime \right]. 
\end{eqnarray}

Using the condition given in Eq. (\ref{expco}) and comparing the coefficients, we obtain the following solutions for $\xi^{'s}$
\begin{eqnarray}
\xi_1&=&\gamma_1\lambda\kappa, \quad\quad \xi_2=-\gamma_2\kappa,\nonumber\\
\xi_3&=&-\gamma_2\kappa,\quad\quad \xi_2=\xi_3.
\end{eqnarray}

The generating functional using Eq. (\ref{jc}) can be expressed as 
\begin{eqnarray}
{\cal{Z}}&=&\int {\cal{D}}\phi^\prime\exp\left\{i\left(S_{eff}+S_1\right)\right\}\nonumber\\
&=&\int{\cal{D}}\phi^\prime \exp\left[i\int d^4x\left\{\Pi^{ia}{\dot{A}}_i^a+\underbrace{\left(1-\gamma_2\kappa\right)}\Pi^{0a}{\dot{A}}_0^a+{\cal{C}}^a{\dot{\cal{P}}}^a +\underbrace{\left(1-\gamma_2\kappa\right)}{\bar{\cal{C}}}^a{\dot{\bar{\cal{P}}}}^a \right.\right.\nonumber\\
&-&\frac{1}{2}\Pi_i^a\Pi_i^a-\frac{1}{4}F_{ij}^aF_{ij}^a+A_0^aD_{ab}^i\Pi_i^b  
+\left.\left.\Pi_0^a\partial_iA^{ia}+\underbrace{\left(\frac{\lambda}{2}+\gamma_1\lambda\kappa\right)}\Pi_0^a\Pi^{0a}+
{\bar{\cal{P}}}^a{\cal{P}}^a \right.\right.\nonumber\\
&-&\left.\left.\partial_i{\bar{\cal{C}}}^aD^i{\cal{C}}^a+gf^{abc}{\cal{P}}^a{\cal{C}}^bA_0^c \right\}\right].
\end{eqnarray}
At $\kappa=0$, the genrating functional will correspond to the usual effective action in Lorentz gauge. At $\kappa=1$, it will correspond to the effective action in axial gauge with the reparameterized gauge parameter $\lambda^\prime=\lambda\left(1+2\gamma_1\right).$ Thus we show that with the appropriate choice of finite BRST parameter in extended phase space, we can  relate the generating functionals corresponding to two different effective theories in different gauges.\\
{\bf Case 2:}\\For another choice of finite BRST parameter $\Theta$ related to $\Theta^\prime$
\begin{equation}
\Theta^\prime=i\gamma\int d^4 y\;{\bar{\cal{C}}}^a(y,\kappa)\;\Pi_0^a(y,\kappa),
\end{equation}
the Jacobian change
\begin{equation}
\frac{1}{J}\frac{dJ}{d\kappa}=i\gamma\int\;d^4y \left[\Pi_0^a(x,\kappa)\right]^2.
\end{equation}
We make a simple ansatz
\begin{equation}
S_1=\xi_1(\kappa)\int d^4 x\left[\Pi_0^a(x,\kappa)\right]^2,
\end{equation}
where $\xi_1(\kappa)$ is a  $\kappa$ dependent arbitrary parameter. Using the condition (\ref{expco}) and comparing the coefficients, we obtain the value  $\xi_1=\gamma\kappa$.
So, the generating functional can be expressed as 
\begin{eqnarray}
Z&=&\int D\chi^\prime\;\exp\left\{i\left(S_{eff}+S_1\right)\right\}\nonumber\\
&=&\int D\chi^\prime \exp i\left[\Pi^{ia}{\dot{A_i}}^a+\Pi^{0a}{\dot{A}}_0^a+{\cal{P}}^a{\dot{\cal{C}}}^a +{\bar{\cal{P}}}^a {\dot{\bar{\cal{C}}}}^a-\frac{1}{2}\Pi_i^a\Pi_i^a-\frac{1}{4}F_{ij}^aF_{ij}^a+A_0^aD_{ab}^i\Pi_i^b \right. \nonumber \\
&+&\left.\Pi_0^a\partial_iA^{ia}+\underbrace{\left(\frac{\lambda}{2}+\gamma\kappa\right)}\Pi_0^a\Pi^{0a}+
{\bar{\cal{P}}}^a{\cal{P}}^a-\partial_i{\bar{\cal{C}}}^aD^i{\cal{C}}^a+gf^{abc}{\cal{P}}^a{\cal{C}}^bA_0^c \right].\label{ficg}
\end{eqnarray}
The generating functional at $\kappa=0$ will be the pure YM theory with a gauge parameter $\lambda$ and at $\kappa=1$, the generating functional will represent pure YM theory with a different gauge parameter $\lambda^\prime=\lambda+ 2\gamma$.

\section{Finite BRST and Canonical Transformations}

The generating functional given in the Eq. (\ref{zfiac}) when integrated over momenta $\Pi_i^a, {\cal{P}}^a, {\bar{\cal{P}}}$ can be expressed in a  compact form 
\begin{equation}
Z=\int D\chi\exp\left\{i\left(S_0+\delta_b\Psi_L\right)\right\},
\end{equation}
where $S_0$ is given by
\begin{equation}
S_0=\int d^4 x \left[\Pi^{ia}{\dot{A_i}}^a-\frac{1}{2}\Pi_i^a\Pi^{ia}+\frac{1}{4}F_{ij}^aF_{ij}^a-A_0^aD^{ab}_i\Pi^{ib} \right],
\end{equation}
and  $\Psi_L$ is the gauge fixed fermion  given by
\begin{equation}
\Psi_L=\int d^4x\left[\frac{\lambda}{2}{\bar{\cal{C}}}^a\Pi_0^a+{\bar{\cal{C}}}^a\partial_iA^{ia}+{\cal{P}}^aA_0^a\right].
\end{equation}
 $\delta_b$ is the BRST variation given by Eq. (\ref{usubrs}).
We now make a canonical transformation in the sector of Lagrange's multiplier which act as a dynamical variables in the BFV formalism as follows
\begin{eqnarray}
\Pi_0^a\quad&\rightarrow&\quad \Pi_0^a-{\dot{A_0^a}},\nonumber\\
A_0^a\quad&\rightarrow &\quad \frac{2}{\lambda}A_0^a.
\end{eqnarray}
They are canonical as they satisfy the following commutation relation for pure YM theory in Lorentz gauge
\begin{equation}
\left[A_0^a({\bf x})\;,\;\Pi_0^b({\bf y})\right]\;=\;i\delta^{ab}\delta^3({\bf x}-{\bf y}).
\end{equation}
Under the above transformations, the Jacobian is unity therefore path integral measure does not change.
The generating functional becomes
\begin{equation}
Z_c=\int D\chi \exp\left\{i\left(S_0+\delta_b\Psi_c\right)\right\}, \label{zfc}
\end{equation}
and the gauge fixed fermion $\psi_L$ changes to $\Psi_C$ in coulomb gauge where 
\begin{equation}
\Psi_c=\int d^4x\left[\frac{\lambda}{2}{\bar{\cal{C}}}^a \Pi_0^a+{\bar{\cal{C}}}^a\partial_iA^{ia}\right].
\end{equation}
The generating functional in Eq. (\ref{ficg}) when integrated over momenta $\Pi_0^a$, ${\cal{P}}^a$, ${\bar{\cal{P}}}^a$ can be expressed in a compact form as given in Eq. (\ref{zfc}).
As shown in the previous section, by choosing a appropriate finite BRST parameter given by Eq. (\ref{tc}), the generating functional corresponding to the effective action in Lorentz gauge $Z_L$ changes to the $Z_C$ i.e. the generating functional corresponding to the effective action in Coulomb gauge.
\begin{equation}
Z_L\stackrel{FFBRST}{---------\longrightarrow}Z_C.
\end{equation}
Thus, we show that the FFBRST transformations are equivalent to the canonical transfromations in the sector of Lagrange's multiplier and its corresponding momenta.
\section{Conclusion}
FFBRST transformations, so far have been studied only in the Lagrangian framework and not in the Hamiltonian formulation. In this paper, we have considered the Hamiltonian formulation of FFBRST transformations using the BFV method. In the BFV Hamiltonian approach, we extend the phase space and the Lagrange multiplier and its corresponding momenta are treated as dynamical variables. In the Hamiltonian formulation, we choose a  BRST parameter in the extended phase space and when integrated over all the momenta, it turns out that this finite parameter only changes the gauge fixed fermion. The generating functional $Z$ does not depend on the choice of gauge fixed fermion. Hence, it plays the same role as in the Lagrangian framework but in a more elegant and simple way. We also show that FFBRST transformations play the role of  canonical transformations for a selective choice of finite BRST parameter.
\\
\\{\bf \Large{Acknowledgment}}\\
We thankfully acknowledge the financial support from the Department of Science and
Technology (DST), Government of India, under the SERC project sanction grant No.
SR/S2/HEP-29/2007. One of us (SKR) would also like to thank CSIR, New Delhi for its financial support.

\end{document}